\documentclass[conference]{IEEEtran}

\usepackage[tight,footnotesize]{subfigure}

\ifCLASSINFOpdf

\else

\fi

\usepackage{graphicx}
\usepackage{cite}
\usepackage{amsmath}
\usepackage{gensymb}
\usepackage{amssymb}

\usepackage{picinpar}
\usepackage{stfloats}
\usepackage{url}
\usepackage{flushend}
\usepackage[latin1]{inputenc}
\usepackage[table]{xcolor}
\usepackage{colortbl}
\usepackage{soul}
\usepackage[caption=false,font=footnotesize]{subfig}
\usepackage{subfloat}

\usepackage{float}
\usepackage{multirow}

\usepackage{upgreek}
\usepackage{textcomp}
\usepackage{balance}

\usepackage[T1]{fontenc}
\usepackage{csquotes}

\makeatletter
\def\ps@IEEEtitlepagestyle{%
	\def\@oddfoot{\mycopyrightnotice}%
	\def\@evenfoot{}%
}
\def\mycopyrightnotice{%
	{\footnotesize \textcolor{white}{978-1-5386-5108-7/17/\$31.00 \textcopyright 2018 IEEE \hfill}}
	\gdef\mycopyrightnotice{}
}

\begin{document}
%


\title{Experimental Body-input Three-stage DC offset Calibration Scheme for Memristive Crossbar}

\author{Charanraj~Mohan\IEEEauthorrefmark{1},
	L.A. Camuñas-Mesa\IEEEauthorrefmark{1},
	Elisa~Vianello\IEEEauthorrefmark{2},
	Carlo~Reita\IEEEauthorrefmark{2},
    José~M.~de~la~Rosa\IEEEmembership{Senior Member,~IEEE,}\IEEEauthorrefmark{1},
	\\Teresa~Serrano-Gotarredona\IEEEmembership{Senior Member,~IEEE,}\IEEEauthorrefmark{1}
	and~Bernabé~Linares-Barranco\IEEEmembership{Fellow,~IEEE}\IEEEauthorrefmark{1}
	\\ \IEEEauthorrefmark{1}Instituto de Microelectrónica de Sevilla, IMSE-CNM (CSIC, Universidad de Sevilla), Sevilla, Spain
	\\ \IEEEauthorrefmark{2}CEA-LETI, Grenoble, France
}

\maketitle

\begin{abstract}
Reading several ReRAMs simultaneously in a neuromorphic circuit increases power consumption and limits scalability. Applying small inference read pulses is a vain attempt when offset voltages of the read-out circuit are decisively more. This paper presents an experimental validation of a three-stage calibration scheme to calibrate the DC offset voltage across the rows of the memristive crossbar. The proposed method is based on biasing the body terminal of one of the differential pair MOSFETs of the buffer through a series of cascaded resistor banks arranged in three stages- coarse, fine and finer stages. The circuit is designed in a 130 nm CMOS technology, where the OxRAM-based binary memristors are built on top of it. A dedicated PCB and other auxiliary boards have been designed for testing the chip. Experimental results validate the presented approach, which is only limited by mismatch and electrical noise.
\end{abstract}

\IEEEpeerreviewmaketitle

\section{Introduction}

\begin{figure}[!tb]
	\centering
	\includegraphics[width=8cm,keepaspectratio=true,height=5cm]{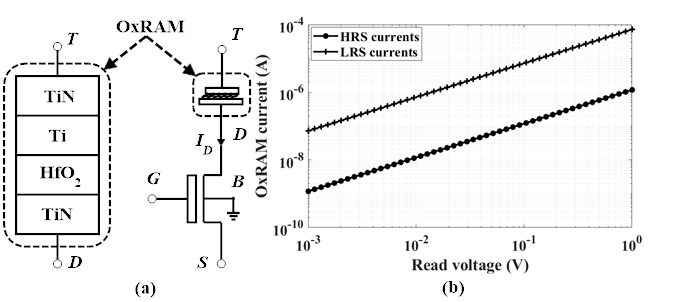}
	\caption{(a) 1T1R memristive synapse structure (b) OxRAM currents for low read voltage pulses.}
	\label{fig:only_1t1r}
\end{figure}

Memristors evolved as a good choice of candidature for artificial synapses in neuromorphic circuits after HP labs proved the physical existence of Chua's finding \cite{Chua71, Chua76, Stru08, Yang08}. Oxide-based Random-Access Memory (OxRAM) slowly rose to one of the promising synapses in neuromorphic computing systems due to their low-switching energy and high endurance \cite{Wase09, Yu11, Kim08, Ha11}. The binary OxRAM synapses in a crossbar architecture used in this work are first formed and then switched between Low Resistance State (LRS) and High Resistance State (HRS) by applying a controlled voltage via a series connected MOSFET \cite{Garb15}. To prevent sneak-path currents, a selector MOSFET is connected in series to the OxRAM, leading to the so called-- \enquote*{1T1R} structure as shown in Fig.~\ref{fig:only_1t1r}(a) \cite{Cass13}.

The filament of the OxRAM is formed by applying a bias $V_{TS}$= 4 V, 10 $\mu$s pulse and gate bias $V_{GS}$= 1 V, with a recommended compliance forming current of about 1 $\mu$A. For a RESET operation, a bias of $V_{ST}$= 3 V, 100 $n$s pulse is applied by keeping the gate fully ON ($V_{GT}$= VDD). For a SET operation, a bias  $V_{TS}$= 2.4 V, 100 $n$s pulse is applied along with the gate bias, $V_{GS}$= 1.5 V. For a read operation, a read voltage of $V_{TS}$ or $V_{Read}$= 0.3 V is applied with a gate bias, $V_{GS}$= 3.8 V. During inference operation, when read pulses are applied across several memristors, power dissipation becomes critical, which limits the scalability of the crossbar \cite{Chicca14}. To overcome this, we need to apply small inference read pulses. Fig.~\ref{fig:only_1t1r}(b) shows OxRAM currents for read voltage pulses less than 1 V when LRS= 13.7 k$\Omega$ and HRS= 845.9 k$\Omega$. However, applying such small read pulses becomes non-trivial, when the offset voltage of the system ruins the measurement. Therefore, the opamps used in the read circuit or the opamps used to buffer crossbar lines should be finely calibrated to keep their DC offset voltages as low as possbile. Conventional calibration schemes exist that compensate offset ranges in the order of few mV \cite{Nagy13, Gines15}. However, this is not enough to increase the required scalability in neuromorphic processors. For this purpose, a three-stage DC offset calibration scheme is proposed by varying the bulk voltage of one of the differential pair MOSFETs of the buffer using a cascaded resistor ladder such that, a calibration step less than 0.1 mV is obtained \cite{Moha18}. This paper experimentally validates the proposed calibration approach. The circuits are designed in a 130 nm CMOS technology on which the OxRAMs are built. The calibration scheme can be applied to crossbars of any size irrespective of the type of synapse. Extensive simulations were carried out during design considering several variations such as technology process corner, monte carlo, temperature, noise and parasitic effects of the layout. A PCB is designed for having full control of the calibration scheme and the experimental results validate the proposed calibration approach. The results are only limited by mismatch, electrical noise and other fabrication defects such as nanobattery effect in OxRAM \cite{Tapper14}.

The rest of the paper is organized as follows: In Section II, the $N\times n$ 1T1R crossbar with calibration of DC offset voltage in each row is described. The three-stage substrate-based calibration scheme is also explained in this section. Section III presents the experimental setup of DC offset calibration scheme. Section IV depicts the experimental results of calibration scheme. Finally, conclusions are drawn in Section V.

\begin{figure*}[!tb]
	\centering
	\includegraphics[width=10.75cm,keepaspectratio=true,height=10.75cm]{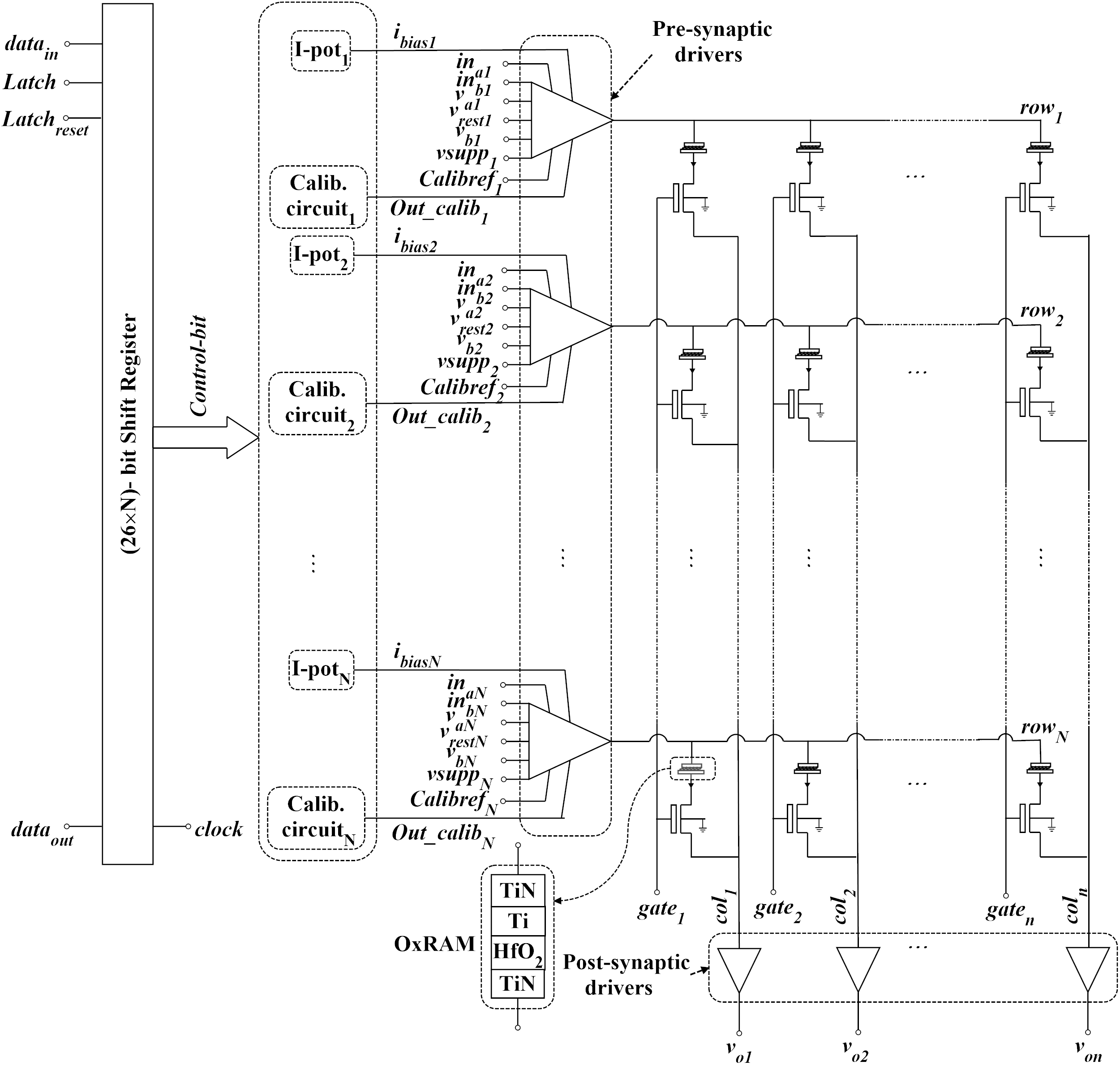}
	\caption{Conceptual schematic of a $N\times n$ 1T1R crossbar with calibration scheme in each row.}
	\label{fig:array}
\end{figure*}

\section{$N\times n$ 1T1R crossbar with calibration of DC offset voltage in each row}

Fig.~\ref{fig:array} shows an $N\times n$ 1T1R memristive crossbar architecture with calibration of DC offset voltage in each row. Each row has its own pre-synaptic driver. Each pre-synaptic driver comprises an opamp, calibration circuit, pulse-shaping digital block and an I-pot. I-pots are digitally programmable current sources which, from a reference current can provide desired current with high precision, down to pA \cite{Serrano07}. I-pots serve as current source biases for the opamps and are controlled by 14-bit control word. Opamps are P-MOSFET based differential two-stage opamps. Each column has its own post-synaptic driver. Each post-synaptic driver comprises an integrator and a comparator. 

\begin{figure}
	\centering
	\includegraphics[width=\columnwidth,keepaspectratio=true,height=5.5cm]{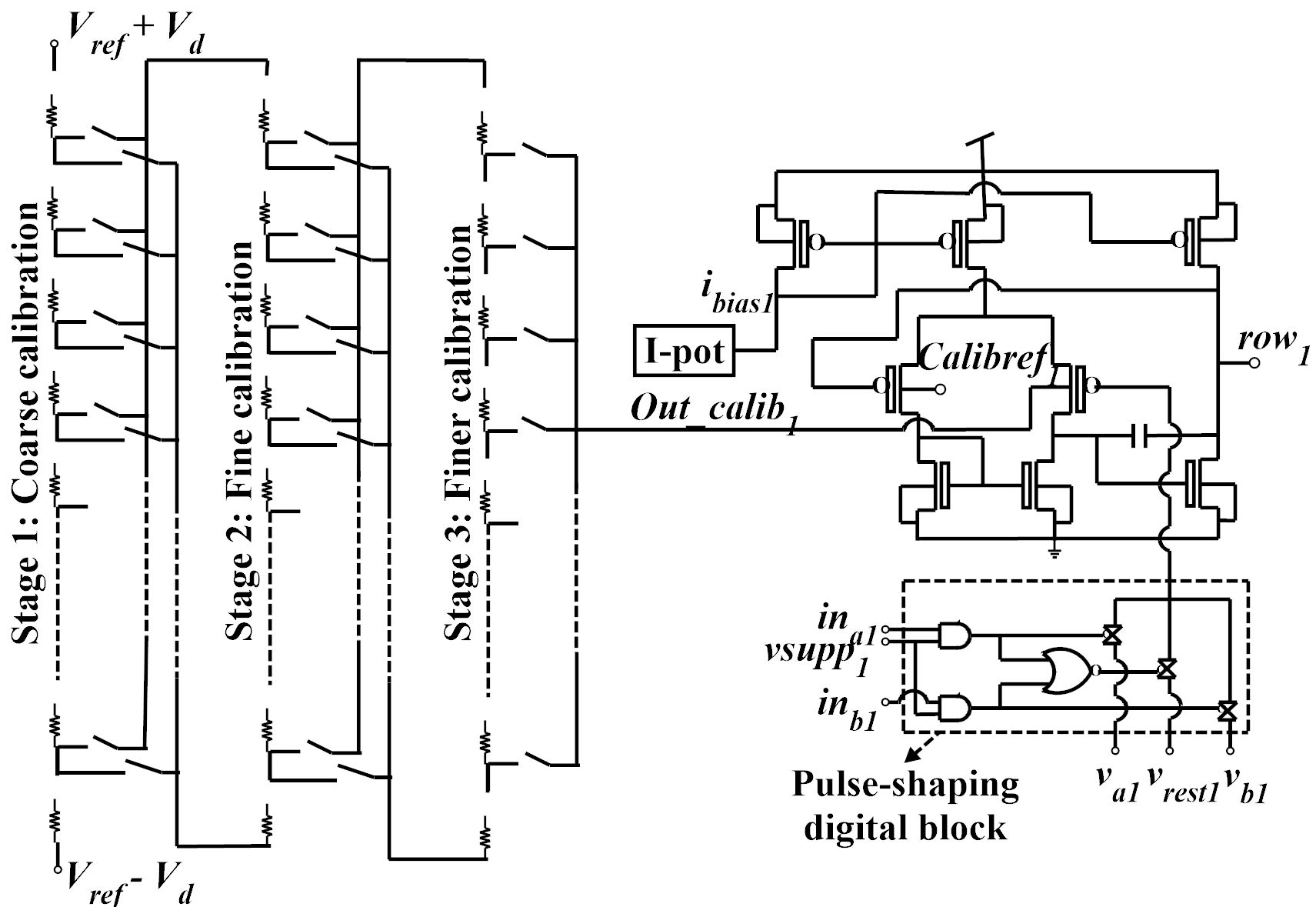}
	\caption{Three-stage calibration scheme of the opamp across $row_{1}$.}
	\label{fig:calibration}
\end{figure}

Fig.~\ref{fig:calibration} shows the three-stage calibration scheme, pulse-shaping digital block and opamp across $row_{1}$. A similar schematic exists in each row. The pulse-shaping digital block is used to set three possible biases through digital control. The three-stage (coarse, fine and finer) calibration scheme is a cascade of resistor ladders whose resistor combinations are chosen by selectively turning ON the P-MOSFET switches via decoders. One of the differential pair MOSFET's body-voltage of the opamp is biased with the calibration range between $V_{ref}$ - $V_{d}$ and $V_{ref}$ + $V_{d}$ in order to compensate the offset across the rows, while the other MOSFET's body-voltage is biased with $Calibref_{1,2,..N}$. $V_{ref}$ is the calibration reference voltage and $V_{d}$ is the calibration differential voltage. The calibration scheme is digitally controlled by a 12-bit control word. To serve this purpose, a 26$\times$N-bit  edge-triggered D-flip flop based shift register is designed for having full-digital control of the I-pots and calibration schemes of the $N\times n$ crossbar.

\begin{figure*}[!tb]
	\centering
	\includegraphics[width=12.6cm,keepaspectratio=true,height=12.6cm]{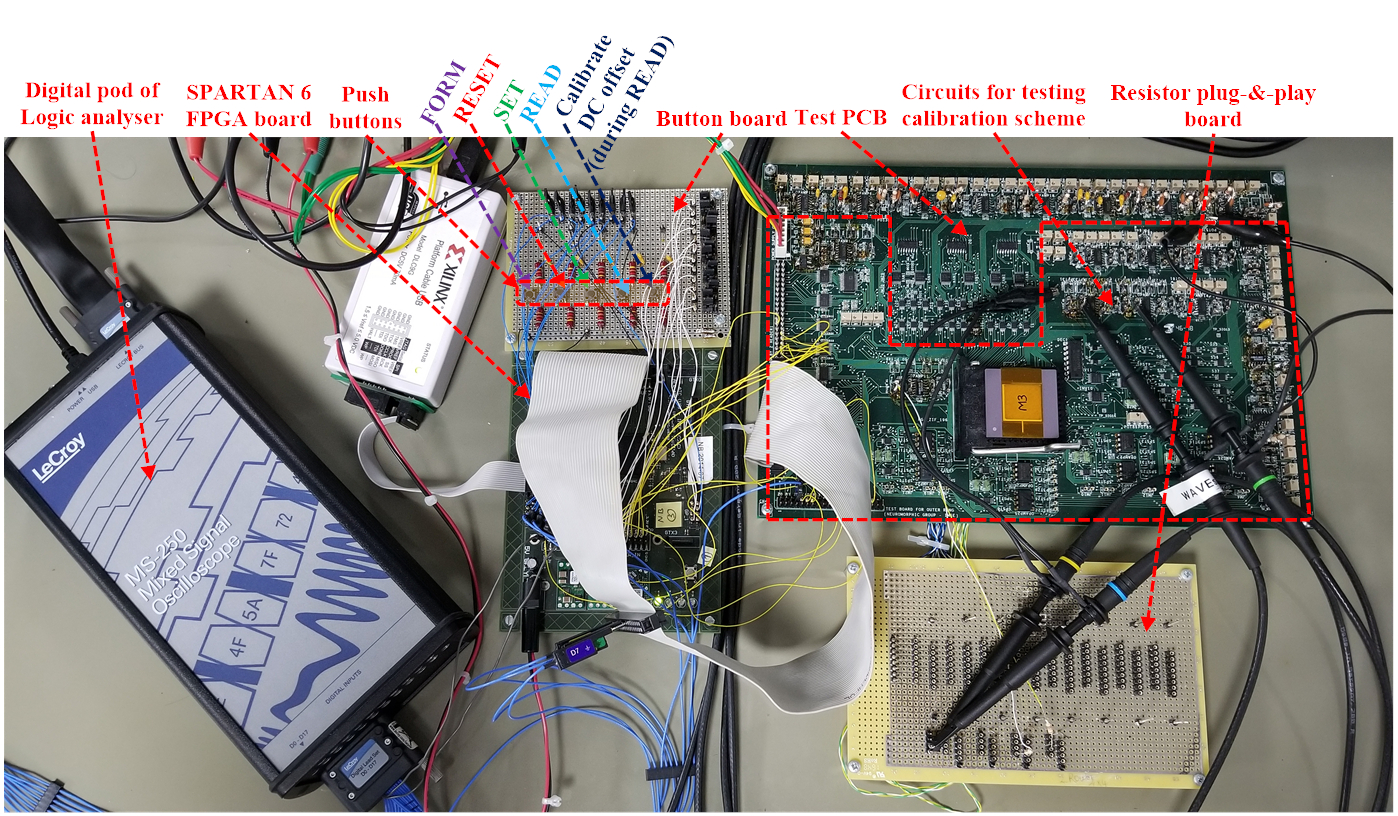}
	\caption{Experimental setup of the DC offset calibration scheme.}
	\label{fig:experi}
\end{figure*}

\begin{figure*}[!tb]
	\centering
	\includegraphics[width=12.6cm,keepaspectratio=true,height=12.6cm]{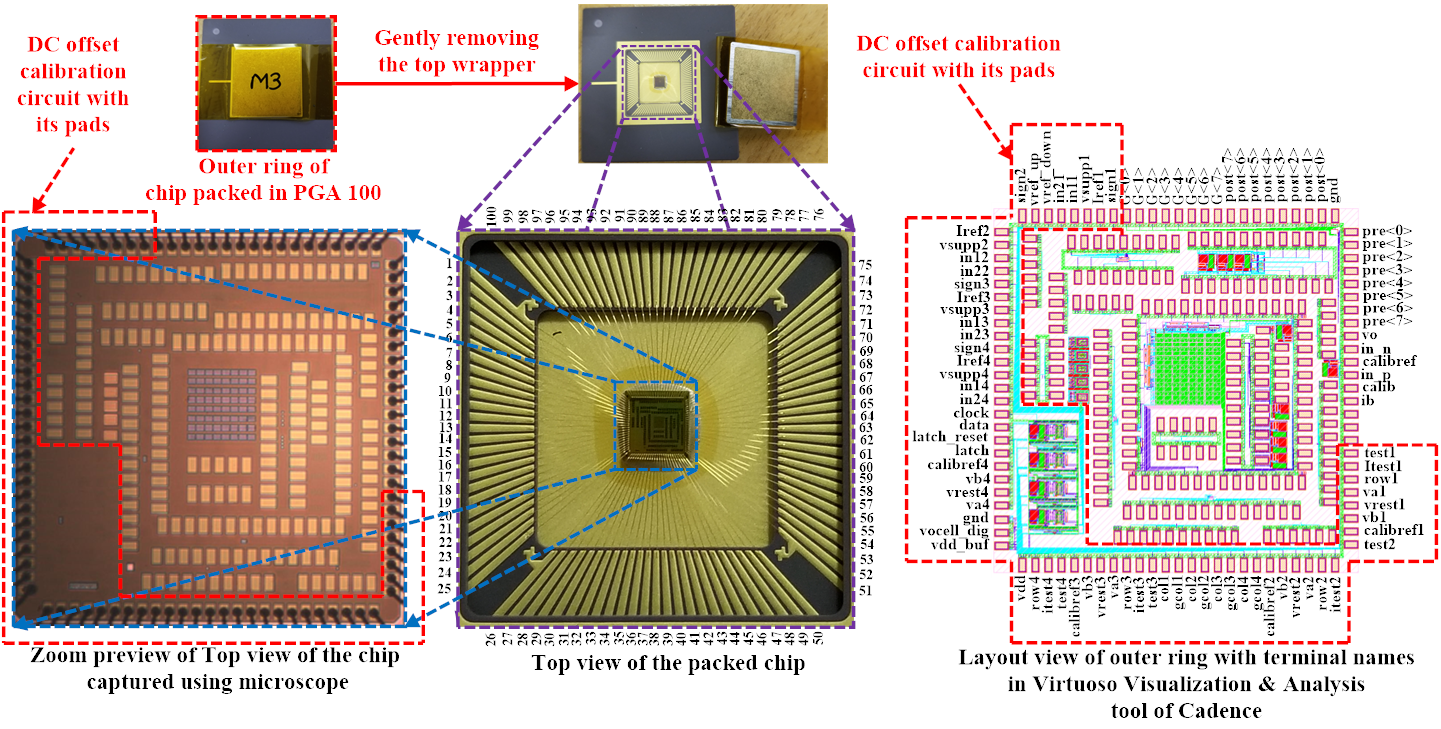}
	\caption{Different previews of the packed chip with its labelled layout.}
	\label{fig:preview}
\end{figure*}

\section{Experimental setup of DC offset calibration scheme}

Fig.~\ref{fig:experi} shows the experimental setup of the DC offset calibration scheme. It mainly comprises the test PCB which includes the chip under test, a SPARTAN 6 driver board, a button board, a resistor plug-and-play board, a logic analyser and its digital pod. Calibration circuit or scheme is part of the circuits in \enquote*{outer-ring} of the chip designed using 130 nm CMOS technology, which has the OxRAMs integrated above it. The \enquote*{outer-ring} of the chip is packed in a 100-pin PGA package, which is connected through a PGA 196-ZIF socket mounted on the test PCB.  Fig.~\ref{fig:preview} shows different previews of the chip packed in PGA100 package with its labelled layout. For testing calibration scheme, a dedicated PCB is made, which mainly comprises components like opamps, level-shifters (3.3 V to 4.8 V), ADC, linear voltage regulator, switches and digital components like decoders, inverters, etc. Opamps on PCB are either used as power supplies opamps or integrating opamps or as comparators. The main purpose of the PCB is to assure desired analog biases at specific terminals of the chip, which are controlled by switches and digital circuits. These switches and digital circuits are further controlled by the SPARTAN 6 driver board. The button board has dedicated buttons to perform OxRAM operations- like FORM, SET, RESET, READ and to calibrate DC offset during READ. It also has jumper arrangements where one can choose the target synapse in a crossbar and also pick the input bit sequence for the calibration scheme.  The button board and the SPARTAN 6 driver board are used together for two main purposes: (i) Target a synaptic 1T1R device in the crossbar and perform operations like FORM, SET, RESET and READ through a 3-bit control signal, ($A$,$B$,$C$) and (ii) Set the 12-bit control word for calibration scheme and perform calibration of DC offset during a READ operation.

\section{Experimental results of calibration scheme}
\begin{figure}
	\centering
	\includegraphics[width=\columnwidth,keepaspectratio=true,height=6.5cm]{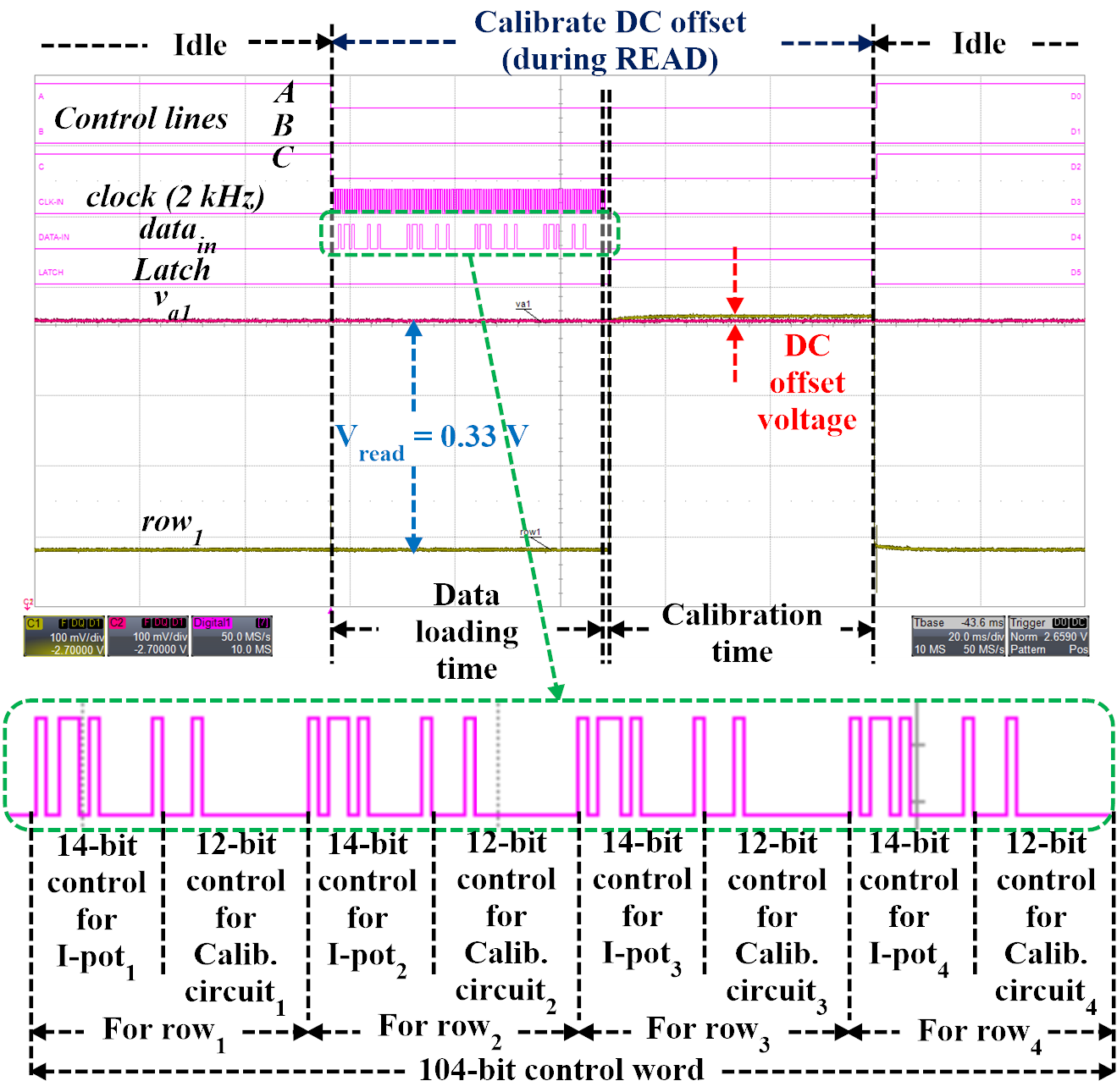}
	\caption{Preview of output screen when calibrating DC offset across $row_{1}$.}
	\label{fig:caliscreen}
\end{figure}

This section shows various results of the three-stage calibration scheme implemented in a $4\times 4$ memristive crossbar. Fig.~\ref{fig:caliscreen} shows a preview of the output screen when $row_{1}$ is calibrated when $V_{read}$= 0.33 V. After forming the targeted OxRAM, a READ operation is performed when $A$= OFF, $B$= OFF and $C$= OFF. Here ($A$,$B$,$C$) is the control signal that is used to set specific OxRAM operation. $data_{in}$ is the 26$\times$4= 104-bit control word, which are the control-bits for I-pots and calibration circuits of the opamps in the experimented $4\times 4$ crossbar. Frequency of $clock$ input to shift-register is kept at 2 kHz. Once, the calibration scheme is biased with $V_{ref}$= 4.5 V, $V_{d}$= 15 mV and $Calibref_{1,2,3,4}$= 4.5 V, $data_{in}$ is loaded into the shift-register. Following this, $Latch$ is turned ON and the targeted row is calibrated by varying the 12-bit input calibration sequence using button-board.

\begin{figure}[!t]
	\centering
	\includegraphics[width=\columnwidth,keepaspectratio=true,height=5.5cm]{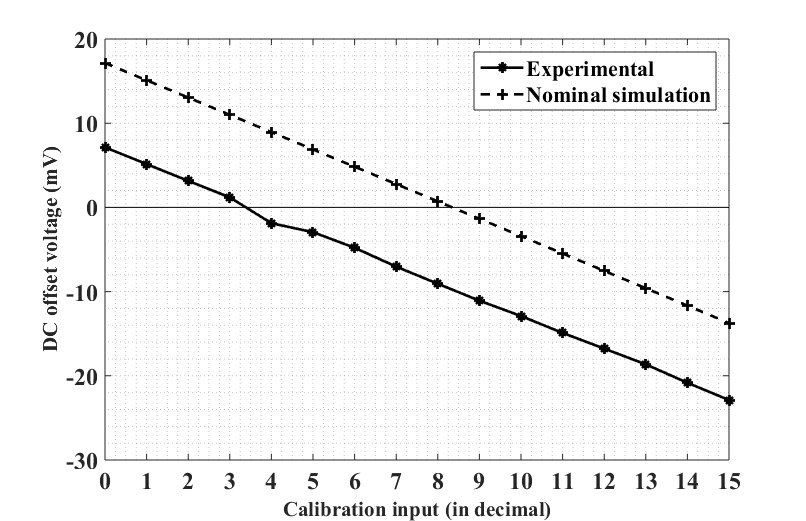}
	\caption{Comparison of experimental and simulation results during stage 1 calibration of DC offset voltage across $row_{1}$.}
	\label{fig:coarse}
\end{figure}

\begin{figure}[!t]
	\centering
	\includegraphics[width=\columnwidth,keepaspectratio=true,height=5.5cm]{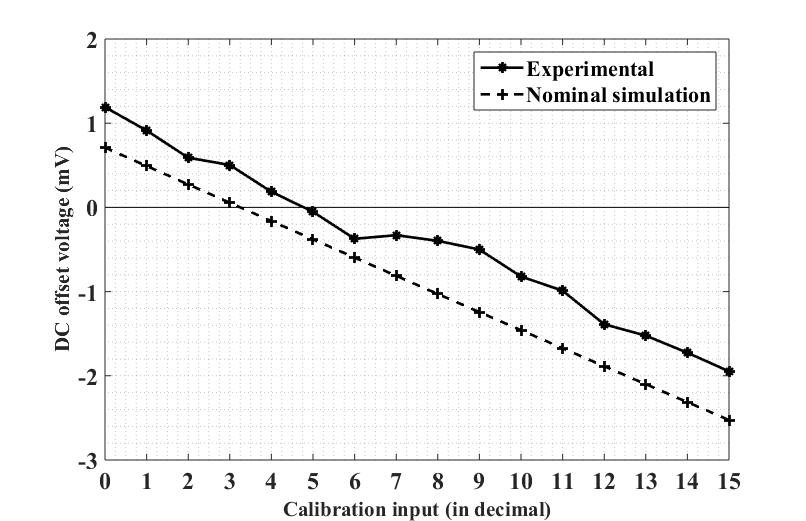}
	\caption{Comparison of experimental and simulation results during stage 2 calibration of DC offset voltage across $row_{1}$ targeting the zero-crossing region.}
	\label{fig:fine}
\end{figure}

\begin{figure}[!t]
	\centering
	\includegraphics[width=\columnwidth,keepaspectratio=true,height=5.5cm]{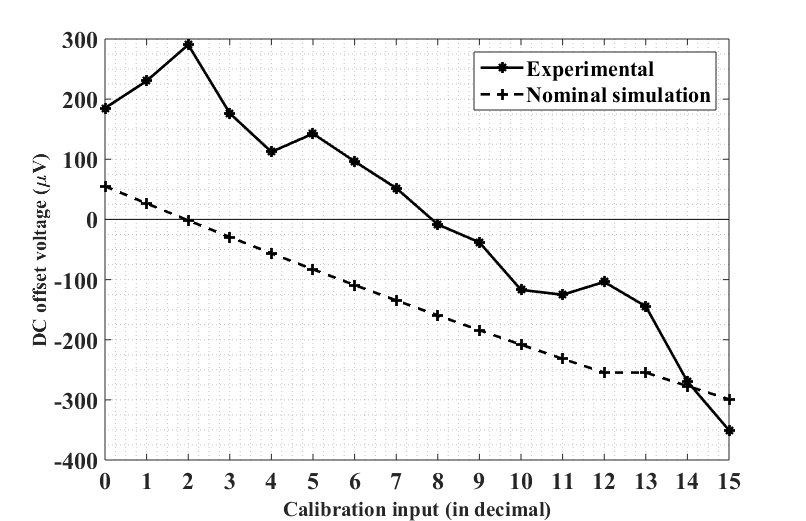}
	\caption{Comparison of experimental and simulation results during stage 3 calibration of DC offset voltage across $row_{1}$ targeting the zero-crossing region.}
	\label{fig:finer}
\end{figure}

Fig.~\ref{fig:coarse} shows the comparison of experimental and simulation results when $row_{1}$ of the crossbar is calibrated for DC offset voltage during stage 1 calibration for $V_{Read}$= 0.33 V. These results are taken by averaging 100 million samples in order to filter out noise, whose standard-deviation is about 200 $\mu$V. The power dissipation during inference READ operation is about 0.8 $\mu$W for a 4$\times$4 crossbar when using a 50 mV read pulse whose DC offset voltage is finely calibrated. The zero-crossing region in Fig.~\ref{fig:coarse} is targeted and DC offset voltages are calibrated during stage 2 and stage 3 calibration, whose results are shown in Fig.~\ref{fig:fine} and Fig.~\ref{fig:finer}. Experimental results of the three-stage calibration scheme match simulation results.

\section{Conclusions}
The experimental results presented in this paper show that it is possible to reduce the DC offset of memristive-array read-out systems below 0.1mV. The proposed approach -- based on the use of a bulk-input differential pair -- is demonstrated by presented measurements, thus opening doors to increasing the scalability of memristive-based neuromorphic systems thanks to the use of lower pulse amplitudes with the subsequent benefits in terms of power dissipation.


\section{Acknowledgements}

This work has been supported in part by the EU H2020 grants 687299 NeuRAM$^3$, 824164 HERMES, 871501 NeurONN, 871371 MeM-Scales, by the Spanish Ministry of Economy and Competitiveness (with support from the European RDF) under contracts TEC2015-63884-C2-1-P (COGNET) and by G0086 ICON. Luis A. Camuñas-Mesa was funded by the VI PPIT through the Universidad de Sevilla.

\end{document}